\documentstyle[12pt]{article}
\advance\topmargin by -5\baselineskip
\advance\textheight by 4\baselineskip
\begin{document}
\title{A novel approach to decays, gluon distributions, and fragmentation
functions of heavy quarkonia}
\vspace{5cm}

\author{Rafia Ali    \\
Department of Physics    \\
Quaid-e-Azam University  \\
Islamabad, Pakistan. \\
\vspace{0.5cm}\\
and    \\
\vspace{0.5cm}\\
Pervez Hoodbhoy   \\
Center for Theoretical Physics  \\
Massachussets Institute of Technology   \\
Cambridge, MA 02139, USA   \\
and  \\
Department of Physics    \\
Quaid-e-Azam University  \\
Islamabad, Pakistan.}
\date{}
\vspace{1cm}

\maketitle
\pagebreak
\begin{figure}
\begin{abstract}
An effective, low-energy, field theory of s-wave quarkonia, constituent
heavy quarks,
and gluons is constructed which is manifestly gauge invariant.
The interaction Lagrangian has the form of a twist expansion,
as typically encountered in hard processes, and involves derivatives
 of arbitrary order.
The parameters in the interaction are related with the non-relativistic
wave function, and standard results for $Q\bar{Q}$ inclusive decays and
radiative transitions are shown to be easily recovered. The light-cone
gluon momentum distribution at very small $x$ is calculated and shown
to be uniquely determined by the
non-relativistic wave function. The distribution has a part
which goes as $x^{-1}logx$, ie. is more singular than the
usually assumed $1/x$ behaviour. The fragmentation function for a virtual
gluon to inclusively decay into an $\eta_c$ or $\eta_b$ is also calculated.
We find that the emission of low momentum gluons makes this process quite
sensitive to assumptions about the binding of heavy quarks in
quarkonia.
\end{abstract}
\end{figure}
\pagebreak

\section{\bf Introduction}

Heavy quarkonium is traditionally modeled as a non-relativistic, colour
singlet bound state of a $Q\,\bar{Q}$   pair with a static coulomb potential
at short relative distances, and some confining type of potential at long
distances.  The binding energy, which is small relative to the heavy quark
masses, is taken as justification for low quark relative velocity as well
as the neglect of explicit gluon degrees of freedom in the hadron's
wavefunction. In hard processes, the entire
non-perturbative QCD  physics is buried into a single parameter: the
wave function at the origin  for the case of S-wave quarkonia, or its
derivative for the case of P-waves. This model has been widely used in
calculating the decay rates of quarkonium states, as well as their production
in
$e^+e^-$ collisions, deep inelastic processes, $Z^0$ decays, etc [1,2].

The problem with this traditional approach to quarkonium modeling is
that gauge invariance - which  is obviously fundamental to QCD because
it is its  legitimising principle - is not respected as an exact symmetry.
Under a local gauge
transformation $q(\vec{x},t) \rightarrow  U(\vec{x},t) q(\vec{x},t)$, the
state normally used to describe quarkonia,  
\newcommand{\be}{\begin{equation}}
\newcommand{\ee}{\end{equation}}
\be
\int d^3 x_1 d^3 x_2 f\left(\vec{x}_1-\vec{x}_2\right)
\bar{q}\left(\vec{x}_1,t\right) \Gamma \,\, q\left(\vec{x}_2, t\right)
\mid 0 >
\ee
does not remain invariant.
In the above equation $\Gamma $ is a space-time independent matrix in spin,
colour, and flavour indices and $f(\vec{x})$  is the relative wave function.
However, one can construct gauge invariant states in the following manner [3]:
define,
\begin{equation}
Q(\vec{x},t) = U^{-1}(\vec{x},t) q(\vec{x},t).
\end{equation}
\begin{equation}
U(\vec{x},t) = P e^{ig \int^{\vec{x}}_{0} d\vec{y} \,\cdot  \,
\vec{A}(\vec{y},t)}.
\end{equation}
Then, the state constructed from $Q$ and $\bar{Q}$,
\begin{equation}
\int d^3 x_1 d^3 x_2 f\left(\vec{x}_1-\vec{x}_2\right)
\bar{Q}\left(\vec{x}_1,t\right) \Gamma \,\, Q\left(\vec{x}_2, t\right)
\mid 0 >
\end{equation}
is indeed invariant for arbitrary $f(\vec{x})$. The gluon field $A^\mu$
transforms in the usual way,
\begin{equation}
A^\mu \rightarrow U^{-1} A^\mu U - \frac{i}{g} U^{-1} \partial^\mu U.
\end{equation}
Unfortunately the operators $Q(\vec{x},t)$ are not pure
(current) quark fields; they also involve arbitrary numbers of soft gluons
which are responsible for  transporting colour between the quarks and
for quarkonium binding. These constituent quarks are clearly extremely
complicated objects. Therefore, to make a gauge invariant model of
quarkonium
requires more than that suggested by Eq.(1).  Fortunately, traditional
quarkonium
models are not too widely off the mark because quarkonia are fairly small and
the path-ordered integral in Eq.(4) is possibly negligible. For p-waves one
expects that
the problem would be more acute  than for s-waves since a centrifugal
barrier serves to keep the quarks apart, leading to a larger meson.
However, to our knowledge, the validity of using a non-gauge invariant state
for either s or p systems has not been investigated.


\section{The Model}

Heavy quarks can be considered as external sources placed in a gluonic
vacuum which undergoes non-perturbative fluctuations, and results in a
modification of the potential type of interaction between quarks in
quarkonium[4].  For large quark mass m, one can consider a $Q\,\bar{Q}$
 pair localized at a relative distance R such that $m \gg 1/R$ and hence
the relative momentum of quarks $p\ll m$. This together with the assumption
that the quarkonium system is weakly bound, ensures that the non-relativistic
approximation   is valid. At the same time, we would like perturbative
methods to be applicable, i.e. $\alpha_s(p) \ll 1$, and hence that $R \ll
\frac{1}{\Lambda_{QCD}} $.   We shall assume that charmonium systems
(bottomonium is obviously better) fulfil the requirement of being
sufficiently heavy, yet also sufficiently  small and weakly bound.

We would like to construct an effective theory of quarkonia, constituent
quarks, and gluons which respects gauge   invariance. Towards this end,
consider   an elementary pseudoscalar meson field $\phi(x)$, representing an
$\eta_c$ or $\eta_b$ meson
for example, which interacts with quark fields $Q(x)$ according to
$\bar{Q} \gamma_5 Q \phi$.    This is clearly wrong since quarkonia are
extended, weakly bound, systems which can be formed only when the
heavy quark   and antiquark happen to have small relative velocities.
In contrast, a point coupling gives an amplitude  for meson formation
independent of relative velocity. To remedy this situation,
and introduce the appropriate non-locality, consider an effective
interaction with an
arbitrary number of    derivatives:
$$-i \sum_{n=0}^{\infty} a_n \bar{Q}(x)\left( \frac{i
\stackrel{\rm \leftrightarrow}{\partial}
 \cdot  \, i \stackrel{\rm \leftrightarrow}{\partial}}{M^2} \right)^n
\gamma_5 Q(x) \phi(x) $$
Here $a_n$ are dimensionless   numbers, to be determined later, and M
is the quarkonium mass.  Before we show that this leads to conventional
formulae for decays etc, we ask what is its gauge-invariant generalization.
This is too unwieldy in general. But as discussed earlier, perturbation
theory     holds in this model and so we can meaningfully consider  a
gauge-invariant model  at the one gluon level. The general form of this
must be:
\begin{eqnarray}
{\cal{L}}_{PS} = &-&i\sum^\infty_{n=0} a_n \bar{Q} \gamma_5 \left(
\frac{i\stackrel{\rm \leftrightarrow}{D} \cdot\, i\stackrel{\rm
\leftrightarrow}{D}}
{M^2}\right)^n Q \phi \nonumber \\ &-& \frac{ig}{2m^2} \sum^\infty_{n=0}
b_{n+1} \bar{Q} \gamma_5 \sigma_{\mu \nu} \left\{ \left(
\frac{i\stackrel{\rm \leftrightarrow}{D} \cdot\,
i\stackrel{\rm \leftrightarrow}{D}}
{M^2}\right)^n F^{\mu \nu} \right\}_{sym} Q \phi .
\end{eqnarray}
where  $\stackrel{\rm \leftrightarrow}{D}^\mu$  is the usual covariant
derivative,
\begin{equation}
\frac{i}{2} \stackrel{\rm \leftrightarrow}{D}^\mu =
\frac{i}{2} \left( \stackrel{\rm \rightarrow}{\partial}^\mu -
\stackrel{\rm \leftarrow}{\partial}^\mu \right) - g
\frac{\lambda_a}{2} A^\mu_a,
\end{equation}
and the symmetrization braces are defined by
\begin{equation}
{\left\{ \;\;\right\}}_{sym} = \left( i \stackrel{\rm \leftrightarrow}{D}
\cdot\,
i\stackrel{\rm \leftrightarrow}{D} \right)^n F^{\mu \nu} +
\left( i\stackrel{\rm \leftrightarrow}{D} \cdot \,i\stackrel{\rm
\leftrightarrow}{D}
\right)^{n-1} F^{\mu \nu}\left(i\stackrel{\rm \leftrightarrow}{D} \cdot \,
i \stackrel{\rm \leftrightarrow}{D} \right) + \cdots +
F^{\mu \nu}\left( i\stackrel{\rm \leftrightarrow}{D} \cdot \,i
\stackrel{\rm \leftrightarrow}{D} \right)^n.
\end{equation}
The interaction in Eqs. 6-8, consisting of an infinite tower of operators
grouped together by symmetry, is, in a sense, a
twist expansion of the type encountered in hard processes. Here the
 ``hard momentum" is the quarkonium mass. Eq.6 is complete at the leading
twist level; other terms added on to it will be sub-dominant. The model
leads in a
straight-forward manner to Feynman vertices\footnote{Eq.(6) contains
derivatives and therefore $H_{PS} \not= - L_{PS}$. The presence of
derivatives leads to additional terms upon quantization.
To illustrate, suppose $ {\cal{L}}
\sim g\bar{\psi} \gamma^\mu \psi \partial_\mu \Phi $. Then the hamiltonian
contains a term  proportional to $g^2 \left( \psi^{\dag} \psi \right)^2$.
Quartic and higher self-couplings can be neglected at the order of
accuracy of our calculations.\\ }.
These are discussed below.

\vspace{1cm}
{\bf A1) \hspace{0.5cm} \underline{PS-Quark Vertex} (Fig.1a).}
The vertex factor for pseudoscalar coupling to quarks is
\begin{equation}
\gamma_5 F(p^2)
\end{equation}
where,
\begin{equation}
F(p^2) = \sum^\infty_{n=0} a_n \left( \frac{p^2}{m^2}\right)^n,
\end{equation}
and $p^\mu = \frac{1}{2} \left(p_2 - p_1 \right)^\mu $ is the relative
momentum.

Do we have any intuition about $ F(p^2) $ ? Since $\left(p_1\!+\!p_2\right)^2
\!=\!M^2$, it follows that $4p^2\!=\!2p^2_1\!+\!2p^2_2\!-\!M^2$ approaches zero
for
$\epsilon\!=\!2m\!-\!M$ approaching zero, i.e. the weak binding limit. It is
therefore reasonable to expect that $F(p^2)$ is steeply peaked
around $p^2\!=\!0$. These explanations are confirmed in the next section,
where it will be shown that $F(p^2)$ can be expressed directly
in terms of the non-relativistic wavefunction of the quarks.

\vspace{1cm}
{\bf A2)\hspace{0.5cm} \underline{PS-Quark-Electric Gluon Vertex} (Fig.1b).}
The vertex factor for coupling to an $E_1$ gluon originates from expanding
out the covariant derivatives in the first term of Eq.(6) and keeping a
single gluon operator   only. Taking the matrix element indicated in Fig.1b
and organizing the term suitably leads to a rather nice and compact form,

\begin{equation}
\frac{2g}{m^2} \,  \gamma_5 \; \varepsilon \cdot \,p \; {\cal{G}}(p,q).
\end{equation}
The function ${\cal{G}}(p,q)$ is most easily expressed in terms of the
dimensionless variables y and z,

\begin{equation}
{\cal{G}}(p,q) = \frac{F(y) - F(z)}{y - z}\;,
\end{equation}
where,
$$y = \frac{\left(p + \frac{1}{2} q\right)^2}{m^2}\; , $$
\begin{equation}
z = \frac{\left(p - \frac{1}{2} q\right)^2}{m^2}\; .
\end{equation}

Note that this vertex is directly expressible in terms of $F(p^2)$
and hence, as we shall see, in terms of non-relativistic quark wave function.
This is a simple
and direct consequence of gauge invariance, and involves no new assumptions.
This is not true for the other gluon vertex discussed below.

\vspace{1cm}
{\bf A3) \hspace{0.5cm} \underline{PS-Quark-Magnetic Gluon Vertex} (Fig.1c).  }
This vertex follows from systematically expanding the second term in Eq.(6),
keeping only one gluon, and taking the matrix element indicated in Fig.1c.
This yields,
\begin{equation}
i \frac{g}{m^2} \,  \gamma_5 \sigma_{\mu\nu}\; \varepsilon^\nu \;
q^\mu \; \tilde{\cal G} (p,q).
\end{equation}
with,
\begin{equation}
\tilde{\cal G}(p,q) = \frac{\tilde{F}(y) - \tilde{F}(z)}{y - z}\;,
\end{equation}
and,
\begin{equation}
\tilde{F}(p^2) = \sum^\infty_{n=1} b_n \left(\frac{p^2}{m^2}\right)^n.
\end{equation}
$\tilde{F}(p^2)$ is in principle different from $F(p^2)$ although one
can expect a similar functional dependence.

 The symmetry of operators will be
different in different mesons. Since we shall deal with $J/\;\psi$
decays, it is useful to consider the generalization
of the pseudoscalar results. All vertices in Fig.1 are immediately
applicable to the $1^{--}$ system by substituting $\gamma_5 \rightarrow
- i \gamma^\mu$ and contracting with the meson polarization vector.
In the limit of large M, the spin-spin interaction   is weak and therefore
$F_{PS}(p^2) =  F_V(p^2)$.

\section{Conventional Limit}

Now that the Feynman vertices for the model have been made explicit,
several calculations can be done straightforwardly. But first, to
understand the physics of $F(p^2)$, consider the lowest order diagram
(Fig.2a)   contributing to the electric form-factor. The contact term
(Fig.2b) involves a higher power of $p^2$ and is therefore neglected.
Imagine that only the quark has electric charge and the antiquark is
uncharged.   The amplitude, to leading order in the quark relative momentum
p and for small photon momentum $q^\mu$, is
\begin{equation}
{\cal A}\!=\!2 e M^2 \varepsilon^* . P \!\! \int\!\!\frac{d^4 p}{(2 \pi)^4}
\frac{F(p^2) F\!\left[(p\!-\!\frac{1}{2}q)^2 \right]}{\left[\!
\left(p\!+\!\frac{1}{2} P\right)^2\!\!-\!m^2\!+\!i
\varepsilon \right] \!\! \left[ \!\left(p\!-\!\frac{1}{2} P\right)^2
\!\!-\!m^2\!+\!i \varepsilon \right] \!\!\left[ \!\left(p\!+\!
\frac{1}{2} P\!-\!q \right)^2\!\!- \!m^2\!+\!i \varepsilon \right]}
\end{equation}
The $p^0$ integral may be performed by keeping only the contributions of
poles in the vicinity of small $p^0$. This yields, for $q^2=0$,
\begin{equation}
{\cal A} = - 2 i e \varepsilon^* . P \int d^3p \left[
\frac{F\left( \vec{p}\;^2 \right)}{(2\pi)^{3/2} M^{1/2} \left(
\varepsilon + \vec{p}\;^2/m \right)}\right]^2 .
\end{equation}
On the other hand, if we consider a charged spinless particle scattering
from an e.m.  field, this has an amplitude equal to $-2ie\varepsilon^*.
P$. This allows us to identify the factor in the brackets in Eq.(18)
with the non-relativistic quark wave function,
\begin{eqnarray}
\frac{F\left(\vec{p}^2\right)}{(2\pi)^{3/2}M^{1/2}\left(\varepsilon +
\vec{p}\;^2/m\right)}&=& \psi(\vec{p})  \nonumber \\
&=& \frac{1}{\sqrt{4\pi}} {\cal R}(p)
\end{eqnarray}

As a consistency check, we calculate the $\eta \rightarrow 2 \gamma$
decay in the model defined by Eq.(6). In the cm frame both photons have
large energy $M/2$ and therefore the contact term in Fig.3b, which is
sharply damped by the form factor ${\cal G}$, does not contribute. Fig.3a
and its crossed version yield for the amplitude:
\begin{equation}
{\cal A} = 2 e^2 M \varepsilon^{\alpha \beta \gamma \delta}
\varepsilon^*_\alpha(q_1)  \varepsilon^*_\beta(q_2)   q_{1 \gamma}
q_{2 \delta} {\cal I}.
\end{equation}
where,
\begin{equation}
{\cal I}\!=\!\int\!\!\frac{d^4p}{(2 \pi)^4}
\frac{F(p^2)}{\left[\!
\left(p\!+\!\frac{1}{2} P\right)^2 \!\!-\!m^2\!+\!i \varepsilon
\right] \!\! \left[ \!\left(p\!-\!\frac{1}{2} P\right)^2 \!\!-
\!m^2\!+\!i \varepsilon \right] \!\!\left[ \!\left(p\!+\!\frac{1}{2} q_2\!-
\!\frac{1}{2}q_1 \right)^2 \!\!- m^2\!+\!i \varepsilon \right]}
\end{equation}

Performing the $p^0$ integration as before, using Eq.(19) to relate
$F(p^2)$ to the wave
function, doing the final phase-space integration, and summing
over colours, one arrives at the standard expression for
$\eta \rightarrow 2 \gamma$ decays,
\begin{equation}
\Gamma_{\eta \rightarrow 2 \gamma} = 12 \alpha^2 \frac{\mid R(0)
\mid^2}{M^2}.
\end{equation}
Only $R(0) $ enters the expression, which is natural enough since the
two quarks annihilate only when very close together (and a check of the
model in Eq.(6)).
Gauge invariance of the quarkonium state does not play a significant
role in this process.

The emission of a soft gluon or photon, as in the M1 transition $J/\psi
\rightarrow \eta_c + \gamma$ (Fig.4), does bring forth the issue of
gauge invariance in an important way because the contact diagrams
(Figs.4b,4c) are unsuppressed. The amplitude for the process is,
\begin{equation}
{\cal A} = - 4 i \frac{e}{M} \varepsilon^{\alpha \beta \gamma \delta}
\varepsilon^*_\alpha(q)  \varepsilon_\beta(P) P_\gamma
q_\delta \left( {\cal I}_{dir} + {\cal I}_{con} \right).
\end{equation}
Here $M, \, P^\mu, \, \varepsilon^\mu (P)$  refer to
the  $J/\psi$. The ``direct", or conventional term, follows from (Fig.4a)
and its crossed version,
\begin{equation}
{\cal I}_{dir}\!=\!i M^2 \!\!\int \!\!\frac{d^4p}{(2 \pi)^4}
\frac{F_{J/\psi}(p) F_\eta(p\!-\!\frac{1}{2}q)}{\left[\!
\left(p\!+\!\frac{1}{2} P \right)^2 \!\!-\!m^2\!+\!i \varepsilon
\right] \!\! \left[\!\left(p\!-\!\frac{1}{2} P \right)^2 \!\!-
\!m^2\!+\!i \varepsilon  \right] \!\!\left[\!\left(p\!+\!\frac{1}{2} P
\!-\!q \right)^2  \!\!-\!m^2\!+\!i \varepsilon \right]}
\end{equation}
Performing the $p^0$  integration and keeping only the contributions
from the poles in the lower half plane near $p^0 \approx 0$ yields,
\begin{eqnarray}
{\cal I}_{dir} &=& M \int \frac{d^3p}{(2 \pi)^3}
\frac{F_{J/\psi}(p) F_\eta(p - \frac{1}{2}q)}{ \left(\varepsilon +
\vec{p}^2 / m \right)^2}   \nonumber \\
&=& \int dr \, r^2 \, e^{\frac{i}{2} \vec{q}.\vec{r}} R_{J/\psi}(r)
R_{\eta}(r).
\end{eqnarray}
In the above, we have kept only the leading order term in the photon
energy $q^0$.   In the dipole approximation, the
exponential factor is unity and if the two mesons had identical wave
functions then one would simply have ${\cal I}_{dir} = 1$.
Because of the hyperfine splitting this deviates from unity, and a
typical (model-dependent) value    \cite{kang:gnus} is ${\cal I}_{dir} =
0.987$. Another
set of model parameters used by the same authors yields 0.984, 0.920.
{}From the amplitude, Eq.(23), the decay width is readily seen to be,
\begin{eqnarray}
\Gamma_{dir}\left( ^{3\!}S_1 \rightarrow \gamma + ^{1\!} S_0 \right) & = &
\frac{4}{3} \alpha e^2 \, \left( \frac{q}{m} \right)^2 \, q \mid {\cal I}
\mid^2 \nonumber \\
& \approx & 2.41 \, keV .
\end{eqnarray}
This differs substantially from the measured width, $1.11 \pm 0.35 keV$.
This is a well-known problem with the usual charmonium model and a wide range
of explanations exist for the factor of 2-3 discrepancy.
These include relativistic corrections, mixing effects, anomalous quark
magnetic moment, etc. References to these may be found in a recent review
by Schuler        \cite{schuler:gnus}.

The contact terms in Figs.4b and 4c, which are required by gauge invariance,
may also be calculated quite straightforwardly,
\begin{equation}
{\cal I}_{con} =  4 i \int \frac{d^4p}{(2 \pi)^4}
\frac{\tilde{\cal G}(p,q) F(p)}{\left[
\left(p - \frac{1}{2} P + \frac{1}{2} q \right)^2 \!\!- m^2
+ i \varepsilon \right] \!\! \left[ \left(p + \frac{1}{2}  P
- \frac{1}{2}q \right)^2 \!\!- m^2 + i \varepsilon
\right] }
\end{equation}
Since this is a correction term, it is adequate to take $F_{J/\psi}
= F_\eta = F$  and ${\cal G}_{J/\psi} = {\cal G}_\eta = {\cal G}$.
The $p^0$  integration gives, keeping only the contribution of nearly
coincident poles near the origin,
\begin{eqnarray}
{\cal I}_{con} &=& - \frac{4}{M^2} \int \frac{d^3p}{(2 \pi)^3}
\frac{\tilde{\cal G}(p,q) F(p)}{\varepsilon + p^2/m} \nonumber \\
&=& - \frac{2 \sqrt{2}}{\pi M^{3/2}} \int dp \, p^2 \, \tilde{\cal G}(p,q)
R(p).
\end{eqnarray}
Since $\mid \! \vec{q} \! \mid \ll  \mid \! \vec{p} \! \mid $, it is
adequate to replace ${\cal G}$ in Eq.(15) by,
\begin{equation}
{\tilde{\cal G}}(p) = - \frac{m^2}{2p} \frac{d\tilde{F}}{dp}.
\end{equation}
At this point one needs to confront the issue: what is $\tilde{F}$
equal to ?  We have seen that F is related to the electric charge
distribution and is expressible in terms of the n.r  wave function (Eq.(19)).
It is possible to show that $\tilde{F}$ is related to the magnetic response
of the system and can, in principle, also be found from a NRQM calculation.
But this is not immediately useful as these calculations
have been done only for static quantities. Instead, we  make the
physically plausible assumption that $\tilde{F}(p) = \xi F(p)$
where $\xi$ is a scale factor. Substituting this into Eqs.(29) and (28)
yields,
\begin{eqnarray}
{\cal I}_{con} &=& -\,\frac{1}{2} \xi M \int^\infty_0 dp \left(\varepsilon +
p^2 / m
\right) R(p) \frac{d}{dp}p R(p) \nonumber \\
&=& \frac{1}{2} \xi \left( 1 - \frac{M \varepsilon}{2} < \frac{1}{p^2} >
\right).
\end{eqnarray}
Any given quarkonium model allows for the calculation of $\varepsilon$
and $<\frac{1}{p^2}>$. For definiteness, assume a gaussian wavefunction of
the type exp$(-p^2/2\beta^2)$, which yields $<1/p^2> = 2/\beta^2$. Fitting
to the $\eta_c$ decay rate gives $\beta^2 \approx 0.413$ Gev$^2$. With
$m_c = 1.65$ Gev, a commonly used value for the charm quark mass, it follows
that ${\cal I}_{con} = - 0.65 \xi$. Now suppose that the entire
(large!) discrepancy between the measured decay width and the
conventionally calculated value can be attributed to the magnetization
term which, as we have argued, symmetry requirements force us to include
in the Lagrangian, Eq.(6). The total decay rate is,
\begin{eqnarray}
\Gamma\left( ^{3\!}S_1 \rightarrow \gamma + ^{1\!} S_0 \right) & = &
\frac{4}{3} \alpha e^2_Q \, \left( \frac{q}{m} \right)^2 \, q \left(
{\cal I}_{dir} + {\cal I}_{con} \right)^2  \nonumber \\
& = & 1.11 \pm 0.35 \,\, \left( for \, J/\psi \rightarrow \gamma
+ \eta_c \right).
\end{eqnarray}
This suggests that $\xi$ is a number around $0.33-0.66$.
With the magnetization term thus determined, one can make physical
predictions for quarkonium processes involving soft photons or gluons.
An application to gluon fragmentation into heavy quarkonium will be
described in the section after next.
\section{Gluonic Distribution}

In the present model it is possible to calculate the
light-cone distribution of low momentum gluons   in a weakly bound
heavy quarkonium system. It will be shown that requiring gauge invariance
of the hadronic state implies the existence of a term which goes as
$x^{-1} \log x$, which is more singular than the  $x^{-1}$
dependence calculated by Brodsky and Schmidt     \cite{brodsky:gnus} using
simple perturbative
arguments for positronium.  An explicit expression for the coefficient of
the $x^{-1} \log x$ term can be provided in terms of the non-relativistic
wave function.  Although calculable, we shall not worry about the $x^{-1}$
terms as this is anyway a theoretical exercise - stable quarkonium targets
unfortunately  do not exist and so gluonic distribution are not directly
measurable.

The starting point  \cite{rali:gnus} is the formula for the gluon momentum
distribution
inside a spinless hadronic target, written as a correlation of
operators on the light cone,
\begin{equation}
G(x) = x \int \frac{d\lambda}{2 \pi} e^{i \lambda x} < P \mid
A^i(0) A^i \left( \lambda n \right)\mid P >.
\end{equation}
Here $k^\mu$ and $n^\mu$ are two null vectors $k^2\!=\!n^2\!=\!0$ with
$k.n\!=\!1$, $k^-\!=\!n^+\!=\!0$, and ${\cal P^\mu}\!=\!k^\mu\!+ \!\frac{1}{2}
M^2 n^\mu$.

Only the transverse components of the gluon field are involved. Inserting
a complete set of states between the two operators, and
limiting the outgoing $X$ to two quarks only, yields,
\begin{equation}
G(x) = x \int [dp_1][dp_2] \delta(x-q \cdot \,n) \mid < p_1 p_2 \mid A^i \mid
P > \mid^2.
\end{equation}
A summation on physical gluon polarizations (i = 1, 2), as well as
colour indices (a = 1, 8), is implicit. The measure [dp] is,
\begin{equation}
[dp] = \frac{dp^+ \, d^2p_\bot}{2p^+ (2 \pi)^3}.
\end{equation}

We now concentrate upon calculating the matrix element in Eq.(33).
If we limit our interest to the terms most singular in x, it turns out that
the term corresponding to radiation from a
quark line (Fig.5b) can be ignored\footnote{This is the diagram which yields
the  $x^{-1}$ behaviour in the work of Brodsky and Schmidt
\cite{brodsky:gnus}.}. The emission
of a magnetic gluon  from the vertex (Fig.5c) is also subdominant - this
follows because the amplitude, Eq.(14),  vanishes as $q \rightarrow 0$.
The dominant contribution comes from the emission of an electric gluon.
Keeping just this term, we have from Eq.(11),
that,
\begin{equation}
<p_1 p_2 \mid A^\mu \mid P > = 2\frac{g}{m^2} \bar{u}(p_1) \gamma_5
v(p_2) \frac{i D^{\mu\nu}}{q^2} p_\nu {\cal G}.
\end{equation}
The light-cone propagator is used in the above,
\begin{equation}
D^{\mu \nu} = - g^{\mu \nu} + \frac{q^\mu n^\nu + q^\nu n^\mu}{q \cdot \,\,n}
\end{equation}
Keeping only the most singular term at small x, as well the lowest order
term in the quark  relative momentum $p^\mu = \frac{1}{2} \left( p_1^\mu
- p_2^\mu \right)$, and performing the polarization sum,
the squared matrix element is calculated to be,
\begin{equation}
\sum \mid < p_1 p_2 \mid A^i \mid P  > \mid^2 = \frac{32 g^2}{m^2 x^2}
\frac{q^2_\bot (p.n)^2}{q^4} {\cal G}^2
\end{equation}
Again for small x,
\begin{equation}
q^2 = - x^2 M^2 - q^2_\bot \, ,
\end{equation}
\begin{equation}
\left[ dp_1 \right] \left[ dp_2 \right] \delta(x - q \cdot \,\,n) = \frac{dp^+
d^2p_\bot d^2K_\bot}{P^+ (2 \pi)^6} \, ,
\end{equation}
where $K_\bot = p_{1 \bot} + p_{2 \bot}$. This yields,
\begin{equation}
G(x) = \frac{16g^2}{m^4} \frac{1}{x P^+} \int \frac{dp^+ d^2p_\bot
d^2K_\bot}{(2 \pi)^6}  \frac{K^2_\bot p^{+2}}{\left( x^2 M^2 + K^2_\bot
\right)^2}  {\cal G}^2
\end{equation}
{}From Eqs.(12-13) ${\cal G}$ is related to the  wave function eq (19) and its
derivative through
\begin{equation}
{\cal G} = \frac{dF}{dt},
\end{equation}
where $t$ is a dimensionless variable,
\begin{eqnarray}
t & = & \frac{\left(p^2 + \frac{1}{4}q^2 \right)}{m^2}  \nonumber \\
& = & - \left( x^2 + \frac{K^2_\bot}{4m^2} + \frac{p^{2}_\bot}{m^2}
+ \frac{2p^{+2}}{m^2} \right)  \nonumber  \\
& = & - \, x^2 - \eta^2.
\end{eqnarray}
Performing the angular integration yields for the gluon distribution,
\begin{equation}
G(x) = -A \frac{\log x}{x} + O\left( \frac{1}{x} \right)\,\, ,
\end{equation}
where A is a positive constant determined by the non-relativistic quark
wave function,
\begin{equation}
A = \frac{2 \alpha}{3 \pi^3} \int^\infty_0 d\eta \, \eta^4 {\cal G}^2(\eta).
\end{equation}
This is the main result of this section. It shows that demanding gauge
invariance of the hadronic state has a profound effect upon the
distribution of low momentum gluons.
\section{Gluon Fragmentation.}

Our final application of the  model developed in this paper
is to calculate the rate of fragmentation of gluons into quarkonia.
Gluon fragmentation refers to the process of converting highly virtual gluons
into hadronic physical states. The calculation is done in two parts. First,
the fragmentation function is calculated at the scale of the heavy quark mass,
and, second, it is evolved perturbatively from
low to high virtualities. If one assumes that perturbative QCD is valid
even at the scale of the charm quark mass,
then a first principles calculation of fragmentation becomes possible.
Recently Braaten and Yuan[9] have performed such a calculation for gluons
fragmenting into $^1\!S_0$ and $^3\!S_1$ heavy mesons.

In the first part of this section we show that the result of the calculation
of Braaten and Yuan     \cite{braaten:gnus} can be exactly replicated by
considering the ``direct"
diagrams (Fig.6a) implied by the model.
The only difference is that our calculation can be performed entirely
in field theoretical language, which is perhaps
an advantage. In the second part,  we show that long wavelength
magnetic gluons emitted by the contact diagram  (Fig.6b)
augment the previous contribution. We remind the reader that, in the present
model, ``long wavelength" nevertheless means a wavelength
sufficiently small for perturbative QCD to be valid: as discussed earlier
there is a hierarchy of scales, $\Lambda_{QCD} \ll q \ll p \ll M$.

The starting point of the calculation is the expression
for the unpolarised gluon fragmentation function into a specific
quarkonium state with momentum  $P^\mu$.
The reference frame is chosen to be the rest frame of the hadron,
\begin{equation}
D(z) = \frac{1}{2} \int \frac{d\lambda}{2\pi} e^{- i \lambda / z} <0\mid
A^i(0) \mid PX><PX\mid A^i(\lambda n) \mid0>.
\end{equation}
The notation here is identical to that in the previous section,
i.e. $k^2\!=\!n^2\!=\!0$, etc. The expression for $D(z)$ follows from
duplicating
the analysis of Jaffe and Ji[10] and replacing for quark operators with gluon
operators. Since $D(z)$ involves only the ``good" components of the gluon
field,
 it is a twist two quantity. A sum on the unobserved states X is implied.
To see more clearly the physical meaning of Eq.(45), put
$ \mid P X > = C^{\dag}(P) \mid X> $, where $C^{\dag}(P)$ creates a meson of
a given type.
Using completeness of the states $\mid X >$ gives,
\begin{equation}
D(z) = \int \left[ dq \right] \delta\left( \frac{q^+}{P^+} -
\frac{1}{z} \right) \frac{<q\mid C^{\dag}(P) C(P) \mid q>}{<q \mid q>}
\end{equation}
This makes apparent that $D(z)$ is essentially the probability of
finding a specific hadron with + component of momentum equal to $zq^+$.
The transverse momentum $q_\bot$ of the incoming gluon is integrated over.

Now consider the production of a $^1 \!S_0$ state from the process in Fig.6a.
After calculating  traces, the amplitude for the process is,
\begin{equation}
< PX \mid A^\mu \mid 0 > = \frac{i D^\mu_{\;\;\nu}}{q^2}
{\cal A}^\nu_{dir}\; ,
\end{equation}
\begin{equation}
{\cal A}^\nu_{dir} = - 8 g^2 m \varepsilon^{\nu \alpha \beta \gamma}
P_\gamma \varepsilon^*_\beta l_\alpha {\cal I}_{dir} \, ,
\end{equation}
\begin{equation}
{\cal I}_{\; dir}\! =\! \int \! \frac{d^4  p}{(2\pi)^4} \frac{F\left( p\!-\!
\frac{1}{2} l \right)}{\left[ \!\!\,\left( p\!+\!\frac{1}{2} q \right)^2 \!\!
-\!m^2 \!+\!i\varepsilon  \right] \!\!\!\left[\!\!\, \left( p\!-\!\frac{1}{2} q
\right)^2\! \!-\!m^2 \!+\!i\varepsilon \right]\!\!\! \left[\!\!\,\left( p\!+\!
\!\frac{1}{2} q \! - \!l \right)^2 \!\!-\!m^2 \!+ \!i\varepsilon \right]}
\end{equation}
The crossed diagram doubles the above value of ${\cal I}$.  Since the
form-factor, which is essentially the wave function,  restricts the relative
quark momentum to small
values, it is apparent from the denominators in Eq.(49) that the dominant
contribution to the integral comes from the region around $q^2 \approx 4m^2$.
Performing the $p^0$ integral as in the previous applications, and using
Eq.(19), yields for the direct amplitude,
\begin{equation}
<PX \mid A^\mu \mid 0 >_{\!\!dir} = \frac{8g^2}{M^{1/2}} \frac{\psi(0)}
{s(s-M^2)} D^{\mu\nu} \varepsilon_{\nu \alpha \beta \gamma} l^\alpha
\varepsilon^{*  \beta} P^\gamma.
\end{equation}
where $s = q^2$ is the mass of the fragmenting gluon, and we have set
$M^2 \approx 4m^2$. The sum over unobserved states in Eq.(45), which
amounts to an integration over the gluon momentum $l$, leads to,
\begin{equation}
D(z) = \frac{1}{6\pi^2} \frac{g^4 \psi^2(0)}{M} \int^\infty_{M^2/z}
\frac{ds}{s^2} \frac{\left[ \left( 1-2z+2z^2 \right) s^2 - 8sm^2z +
16m^4 \right]}{\left( s - M^2 \right)^2}
\end{equation}
The lower limit of integration follows from setting the minimum
value of $q^2_\bot = l^2_\bot = 0 $ in,
\begin{equation}
s = \frac{M^2}{z} + \frac{z}{1-z} q^2_\bot  \,\, .
\end{equation}
A colour factor of 1/12 has been included in Eq.(51). Performing
the integration yields,
\begin{equation}
D(z) = \frac{\alpha^2_s}{3\pi} \frac{{\mid R(0)\mid }^2}{M^3}
\left[ 3z - 2z^2 + 2(1-z) \log(1-z) \right] .
\end{equation}
This is precisely the result of Braaten and Yuan     \cite{braaten:gnus}, ie.
Eq.(8) of
their paper.


The contact diagram of Fig.6b is calculated similarly and it has the Lorentz
structure given in Eq.(48) with,

\begin{equation}
{\cal I}_{\; con}\! =\! \frac{2}{M^2} \int \! \frac{d^4  p}{(2\pi)^4}
\frac{\tilde{\cal G}(p,l)}{\left[ \!\!\,\left( p+\frac{1}{2} q \right)^2 \!\!
- m^2 \!+ i\varepsilon  \right] \!\!\!\left[\!\!\, \left( p - \frac{1}{2} q
\right)^2\! \!- m^2 \!+ i\varepsilon \right]}.
\end{equation}
Only the vertex Eq.(14-16) is involved; electric gluons do not contribute
here. Since the magnetic form factor $\tilde{\cal G}$ restricts $p$ to small
values, from the two denominators in the above integral it is evident that
the major contribution comes from small values of the outgoing gluon
momentum $l$.  Performing the $p^0$ integration gives,
\begin{equation}
<PX \mid A^\mu \mid 0 >_{\!\!con}
= - \frac{8g^2}{M^3 s \mid \! \vec{l} \! \mid} D^{\mu\nu} \varepsilon_{\nu
\alpha \beta \gamma} l^\alpha \varepsilon^{*  \beta} P^\gamma \int
\! \frac{d^3  p}{(2\pi)^3}  \tilde{\cal G}
\end{equation}
Since we have chosen our reference frame as the rest-frame of the produced
meson, it follows that,
\begin{equation}
\mid \!\vec{l}\!\mid = \left(s-M^2\right)/2M
\end{equation}
Using Eq.(29) yields,
\begin{eqnarray}
\int  \! \frac{d^3  p}{(2\pi)^3}  \tilde{\cal G}(p) &=&
- \frac{m^2}{2}\int  \! \frac{d^3  p}{(2\pi)^3}  \frac{1}{p}
\frac{d\tilde{F}}{dp}. \nonumber\\
&=& \frac{1}{4} M^3 \left(\frac{\tilde{\psi}(0)}{M^{3/2}} +
\frac{\varepsilon}{2M} M^{1/2} \int^\infty_0 dr \, r \,\tilde{\psi}(r)
\right).
\end{eqnarray}
Thus the total amplitude is,
\begin{eqnarray}
&<&PX \mid A^\mu \mid 0 >_{\!\!dir} + <PX \mid A^\mu \mid 0 >_{\!\!con}
\nonumber \\
&=& \frac{8e^2}{M^{1/2}} \frac{1}{s(s-M^2)} D^{\mu\nu} \varepsilon_{\nu
\alpha \beta \gamma} l^\alpha \varepsilon^{*  \beta} P^\gamma
\left( \psi(0) - \frac{1}{2} \chi(0) \right),
\end{eqnarray}
where,
\begin{equation}
\chi(0) = \tilde{\psi}(0) + \frac{\varepsilon M}{2} \int^\infty_0
dr \, r \tilde{\psi}(r).
\end{equation}

Using Eqs.(58-59) to calculate the fragmentation function yields,
\begin{equation}
D(z) = \frac{\alpha^2_s}{3\pi M^3} \eta(z)
\left(R(0) - \frac{1}{2} S(0) \right)^2,
\end{equation}
where,
\begin{equation}
\eta(z) =  3z - 2z^2 + 2(1-z) \log(1-z)
\end{equation}
and
\begin{equation}
S(0) = \sqrt{4\pi} \chi(0).
\end{equation}
The first term in Eq.(60) is the direct term and is the same as Eq.(53),
while the second is the contact term. To numerically estimate $D(z)$, we
shall make the same assumption as in section 3, i.e. that
$\tilde{F}(p)=\xi F(p)$ and use the same gaussian wave function. $D(z)$ can
then be expressed as,

\begin{equation}
D(z) = \frac{\alpha^2_s}{3\pi M^3} \eta(z) R^2(0) \left( 1 - a \right)^2,
\end{equation}
where $a$ varies from 0.35 to 0.71 as  $\xi$ goes from 0.33  to 0.66,
the range estimated in section 3. It is clear from Eq.(63) that except
for a scale factor $(1\!-\!a)^2$ the fragmentation function at the initial
scale,
as well as after evolution, is identical
to that calculated by Braaten and Yuan     \cite{braaten:gnus}.
The scale factor, however, causes
a substantial decrease
in the magnitude of $D(z)$.

\section{Summary}

We have presented in this paper a low-energy, effective, gauge-invariant
theory wherein the fundamental degrees of freedom are quarkonia, quarks,
and gluons. The interaction has the form of a twist expansion familiar
from hard processes and consists of towers of operators grouped together
according to their symmetry. The
quarkonium mass plays the role of the ``hard momentum". The arbitrary number
of derivatives in the theory serve to bring in the appropriate amount of
non-locality or, equivalently, a form-factor in momentum space which embodies
the extended structure of the meson. This form factor, which is the basic
input into the model, was shown to be directly related to the
non-relativistic
quark wavefunction, a quantity calculable in any given potential model.

An important consequence of gauge-invariance is the emergence of Feynman
vertices representing the direct gluon-quark-antiquark-meson interaction.
These vertices have a substantial effect upon certain heavy-meson phenomena.
For example, the radiative $M_1$ transition,
$J/\psi \rightarrow \eta_c + \gamma$ receives an additional contribution from
one such vertex. This could help explain why the usual decay calculations
invariably overestimate the decay rate by a factor of 2-3. As another example,
we have calculated the light-cone momentum distribution of gluons in heavy
quarkonia. Although these distributions are probably of no practical interest,
nevertheless the present model does have some interesting theoretical
consequences. We find that the $Q\bar{Q}G \,\, ^{1\!}S_0$ vertex, which
contributes to $G(x)$, not
only gives the $x^{-1}$ behaviour but also has a $ x^{-1}logx$ part which is
more singular at small x.

As the final application of our model we considered the fragmentation of gluons
into $^{1\!}S_0$ mesons. If we ignore the contact (gauge) diagrams, then the
results of Braaten and Yuan     \cite{braaten:gnus}  are exactly recovered. But
including these
diagrams leads to a downwards rescaling of the their results by an amount
which could be substantial. A proper calculation depends upon knowledge of
the magnetic formfactor, called $\tilde{F}(p^2)$ here, which in principle
could be determined from a quark model calculation that includes states of
arbitrary excitation. Finally, we remark that the model discussed in this
paper is extendable to p states as well. This would be interesting because
the centrifugal barrier keeps the quarks relatively further apart and thus
makes the issue of non gauge-invariance of the meson state more acute.

\vspace{6cm}
\section{Acknowledgements}

We thank John Ellis, Robert Jaffe, Xiangdong Ji, and John Ralston for
discussions. R.A. gratefully acknowledges financial support from Professor
Abdus Salam for her graduate studies. This work was supported in part by
NSF Grant INT-9122027.

\pagebreak

\newpage
{\bf{\Large Figure Captions}}
\begin{enumerate}
\item Vertex factors for the coupling of pseudoscalar quarkonium to quarks
and gluons. a) Quarkonium - quarks coupling; b) and c) coupling to electric
and magnetic gluons respectively. \\
\item Contributions from (a) direct and (b) contact terms to the form factor of
$^1S_0$.\\
\item Contributions from (a) direct and (b) contact terms to the
$\eta \rightarrow 2 \gamma$ decay rate. \\
\item Contributions to $^{3\!}S_1 \rightarrow \gamma + ^{1\!} S_0$ decay
rate from: a) direct diagram; b) contact diagram with $\gamma$ emanating
from $^3S_1$ vertex; c) contact diagram with $\gamma$ emanating from $^1S_0$
vertex. \\
\item Probing the gluon distribution in a $^1S_0$ meson: a) gluon and
unobserved final state hadrons $X$; b) gluon emission from a quark line;
c) gluon emission from $^1S_0$ vertex.\\
\item Gluon fragmenting into $\eta_c(^1S_0)$. a) Direct diagram, and,
b) contact diagram .\\
\end{enumerate}

Figures may be obtained by writing to\\
                    hafsa\%png-qau\%sdnpk@sdnhq.undp.org

\end{document}